\documentclass[12pt]{article}
\usepackage{amsfonts}
\usepackage{amsmath}
\usepackage{amsthm}
\usepackage{amssymb}
\usepackage{mathrsfs}
\usepackage{graphicx}
\usepackage[bf,center]{titlesec}

\renewcommand{\iint}{\int \!\!\!\! \int}

\renewcommand{\iiiint}{\int \!\!\!\! \int \!\!\!\! \int \!\!\!\! \int}
\title{Optimal $\mathfrak{L}^{\beta}$-Control for the Global Cauchy Problem of the Relativistic Vlasov-Poisson System}
\author{Brent Young \\ Rutgers University \\ bojy@math.rutgers.edu}

\begin{document}

\maketitle

\begin{abstract}
Recently, M.K.-H. Kiessling and A.S. Tahvildar-Zadeh proved that a unique global classical solution to the relativistic Vlasov-Poisson system exists whenever the positive, integrable initial datum is spherically symmetric, compactly supported in momentum space, vanishes on characteristics with vanishing angular momentum, and for $\beta \ge 3/2$ has $\mathfrak{L}^{\beta}$-norm strictly below a positive, critical value $\mathcal{C}_{\beta}$. Everything else being
equal, data leading to finite time blow-up can be found with $\mathfrak{L}^{\beta}$-norm surpassing $\mathcal{C}_{\beta}$ for any $\beta >1$, with $\mathcal{C}_{\beta}>0$ if and only if $\beta\geq 3/2$. In their paper, the critical value for $\beta = {3}/{2}$ is calculated explicitly while the value for all other $\beta$ is merely characterized as the infimum of a functional over an appropriate function space.  In this work, the existence of minimizers is established, and the exact expression of $\mathcal{C}_{\beta}$ is calculated in terms of the famous Lane-Emden functions.  Numerical computations of the $\mathcal{C}_{\beta}$ are presented along with some elementary asymptotics near the critical exponent ${3}/{2}$.
\end{abstract}

\section{Introduction}

The relativistic Vlasov-Poisson (rVP) system is given by $$\textrm{rVP}^{\pm}:\;\left\{ \begin{array}{r}\left(\partial_t + \frac{p}{\sqrt{1+\lvert p \rvert^2}} \cdot \nabla_q \pm \nabla_q\varphi_t(q) \cdot \nabla_p  \right)f_t(p,q)=0\\ \\ \triangle_q\varphi_t(q) = 4\pi \int f_t(p,q)\; d^3p \\ \\\varphi_t(q) \asymp -\lvert q \rvert^{-1} \textrm{ as } \lvert q \rvert \to \infty ; \end{array}\right.$$ $\textrm{rVP}^{+}$ models a system with repulsive interaction while $\textrm{rVP}^{-}$ models a system with attractive interaction.  One of the earliest papers to appear on the subject is \cite{GS85}  wherein Glassey and Schaeffer show that global classical solutions to $\textrm{rVP}^{\pm}$ will exist for initial data that are spherically symmetric, compactly supported in momentum space, vanish on characteristics with vanishing angular momentum, and have $\mathfrak{L}^{\infty}$-norm below a critical constant $\mathcal{C}_{\infty}^{\pm}$ with $\mathcal{C}_{\infty}^{+} = \infty$ and $\mathcal{C}_{\infty}^{-} < \infty.$  More recently, Had\v zi\'c and Rein (\cite{HR07}) showed the non-linear stability of a wide class of steady-state solutions of $\textrm{rVP}^-$ against certain allowable perturbations utilizing energy-Casimir functionals.  Shortly thereafter, Lemou, M\'ehats, and Rapha\"el (\cite{LMR08a,LMR09}) investigated non-linear stability and the formation of singularities in $\textrm{rVP}^-$ through concentration compactness techniques.

In this work, we focus exclusively on the attractive case and henceforth suppress the superscript on both $\textrm{rVP}^-$ and $\mathcal{C}_{\infty}^{-}.$  In \cite{KTZ08}, Kiessling and Tahvildar-Zadeh prove that a unique global classical solution to the relativistic Vlasov-Poisson system exists whenever the positive, integrable initial datum $f_0$ is spherically symmetric, compactly supported in momentum space, vanishes on characteristics with vanishing angular momentum, and has $\mathfrak{L}^{\beta}$-norm below a critical constant $\mathcal{C}_{\beta}$, with $\mathcal{C}_{\beta} > 0 $ if and only if
$\beta \ge 3/2$.  The constant $\mathcal{C}_{\beta}$ is critical in the sense that, everything else being equal, initial data can be found which lead to blow-up in finite time if their $\mathfrak{L}^{\beta}$-norm is allowed to be ever so slightly bigger than $\mathcal{C}_{\beta}$.  This critical constant is given by the following minimization problem:
\begin{eqnarray}
\mathcal{C}_{\beta} &\equiv &\inf_{\mathfrak{P}_1 \cap \mathfrak{L}^{\beta}} \Phi_{\beta}(f),\\
\Phi_{\beta}(f) &\equiv&\left( \frac{\mathcal{E}_p^u(f)}{-\mathcal{E}_q(f)}\right)^{3(1-\frac{1}{\beta})}\lVert f \rVert_{\beta},
\end{eqnarray}
where $\mathfrak{P}_1 \cap \mathfrak{L}^{\beta}$ denotes the set of probability measures on $\mathbb{R}^6$ with finite first moment which are absolutely continuous with respect to Lebesgue measure having density in $\mathfrak{L}^{\beta}$.  The functionals $\mathcal{E}_p^u$ and $\mathcal{E}_q$ are given by
\begin{eqnarray}
\mathcal{E}_p^u(f) &\equiv& \iint \lvert p \rvert f(p,q)  \;d^3p \; d^3q,\\
\mathcal{E}_q(f) &\equiv&-\frac{1}{2} \iiiint \frac{f(p',q')f(p,q)}{\lvert q - q' \rvert} \;d^3p' \; d^3p \; d^3q' \; d^3q.
\end{eqnarray}
It is further shown that \begin{equation}\left[ \left(\frac{3}{8} \right)^3 \frac{15}{16} \right]^{1-\frac{1}{\beta}} \le \mathcal{C}_{\beta} \le \frac{45}{8\pi^2}\left(\frac{8\pi^{\frac{5}{2}}}{\prod_{k=1}^{3}(k +2\beta)} \frac{\Gamma(\beta)}{\Gamma\left(\beta+\frac{3}{2}\right)} \right)^{\frac{1}{\beta}}. \label{lb_ub}\end{equation}
In particular, the optimal constant for $\beta = \frac{3}{2}$ is explicitly calculated in the paper.  Its value is $\frac{3}{8}\left( \frac{15}{16} \right)^{\frac{1}{3}}$, which is equal to the one given by either bound.

In this paper, we address the determination of the critical norm for all remaining cases.  We begin in the next section by proving the existence of minimizers for $\Phi_{\beta}$ in a slightly larger class of functions $\Omega_{\beta}$ to be defined below.  After that, we characterize the minimizers via variational techniques and find that they are the well known Lane-Emden polytropes.  This will show that the minimizers are actually in our original space $\mathfrak{P}_1 \cap \mathfrak{L}^{\beta}$.  Finally, we compute the optimal constant in terms of the parameter $\beta$ and the first zero and corresponding slope of the standard polytropes.  At the end, we mention some numerical results pertaining to the calculation of $\mathcal{C}_{\beta}.$

\section{Existence of Minimizers}

To begin with, as remarked in the ``note added" in \cite{KTZ08}, the variational problem (1),(2) is equivalent after rescaling to the one given by Lemou, M\'ehats, and Rapha\"el  in \cite{LMR08b}, who designed it for different purposes, namely to study blow up dynamics for the relativistic Vlasov-Poisson system.  Instead of giving a detailed analysis, they refer the reader to their earlier work \cite{LMR08a} where they study an analogous variational principle (given in formula 1.18) for the non-relativistic Vlasov-Poisson system. Thus, in principle we could build on their results to evaluate $C_\beta$. Instead, here we first give a somewhat different proof of the existence and characterization of minimizers by combining the techniques of Weinstein (\cite{W83}) (also referred to by Lemou, M\'ehats, and Rapha\"el) with those of Lieb and Simon (\cite{LS77}). This strategy has certain advantages, as we shall see in the next section.

We find it convenient (for reasons explained below) to expand the class of functions over which we attempt to minimize $\Phi_{\beta}$.  To that end, define $$\Omega_{\beta} = \{f:\mathbb{R}^6 \to \mathbb{R}: f \ge 0,  \lVert f \rVert_1 + \lVert \lvert p \rvert f \rVert_1 + \lVert f \rVert_{\beta} < \infty \}  $$ and note that $\mathfrak{P}_1 \cap \mathfrak{L}^{\beta} \subset \Omega_{\beta}$ (indeed, a function $f \in \Omega_{\beta}$ will also be in $\mathfrak{P}_1 \cap \mathfrak{L}^{\beta}$ whenever $\lVert \lvert q \rvert f \rVert_1 $ is finite and the $\mathfrak{L}^1$-norm is equal to 1).  Since we have allowed functions of arbitrary $\mathfrak{L}^1$-norm into our considerations, we need to adjust our definition of $\Phi_{\beta}$.  Assuming $\lVert f \rVert_1 > 0$ and inserting $\frac{f}{\lVert f \rVert_1}$ into $\Phi_{\beta}$, we arrive at the appropriate functional: \begin{equation} \widetilde{\Phi}_{\beta}(f) \equiv\left( \frac{\mathcal{E}_p^u(f)}{-\mathcal{E}_q(f)}\right)^{3(1-\frac{1}{\beta})}\lVert f \rVert_{\beta}\;\lVert f \rVert_1^{2-\frac{3}{\beta}}. \end{equation}  We now seek to minimize $\widetilde{\Phi}_{\beta}$ over this enlarged space of functions.  In the next section, we will show that there exist minimizers for $\widetilde{\Phi}_{\beta}$ over $\Omega_{\beta}$ that are also in $\mathfrak{P}_1 \cap \mathfrak{L}^{\beta}$.

The demonstration that minimizers exist closely follows Weinstein (\cite{W83}). Since $\widetilde{\Phi}_{\beta}(f) \ge 0$ for all functions in $\Omega_{\beta}$, we can find a minimizing sequence $\{ f_{\beta,n}\} \subset \Omega_{\beta}$ so that \begin{equation}\widetilde{\mathcal{C}}_{\beta} \equiv \inf_{f \in \Omega_{\beta}} \widetilde{\Phi}_{\beta}(f) = \lim_{n \to \infty} \widetilde{\Phi}_{\beta}(f_{\beta, n}).  \end{equation}  Since we are minimizing over a larger class of functions, $\widetilde{\mathcal{C}}_{\beta} \le \mathcal{C}_{\beta}$ and so the upper bound noted above still holds.  Applying the well-known procedure of spherically symmetric equi-measurable rearrangements (c.f. \cite[Chapter 3]{LL01}), we can assume $f_{\beta, n}(p,q) = f_{\beta, n}(\lvert p \rvert,\lvert q \rvert, \theta)$ (committing a slight abuse of notation) where $\theta$ is the angle between $p$ and $q$.

For any positive real numbers $\kappa,\lambda$ and $\mu$, the triple family of scaling \begin{equation}f_{\kappa,\lambda,\mu}(p,q) \equiv \mu \;f(\lambda p, \kappa q)  \label{scaling}\end{equation} leaves $\widetilde{\Phi}_{\beta}$ invariant while
\begin{eqnarray}
\mathcal{E}_p^u(f_{\kappa,\lambda,\mu}) &=& \frac{\mu}{\lambda^4 \kappa^3}\mathcal{E}_p^u(f),\\
\mathcal{E}_q(f_{\kappa,\lambda,\mu}) &=& \frac{\mu^2}{\lambda^6 \kappa^5}\mathcal{E}_q(f),\\
\lVert f_{\kappa,\lambda,\mu} \rVert_r &=& \frac{\mu}{(\kappa \lambda)^{\frac{3}{r}}}\lVert f \rVert_r.
\end{eqnarray}
Taking advantage of this scaling invariance, we can assume that our minimizing sequence has the following properties:
\begin{eqnarray}
f_{\beta, n}(p,q) &=& f_{\beta, n}(\lvert p \rvert,\lvert q \rvert, \theta),\\
\mathcal{E}_p^u(f_{\beta, n}) &=& 1,\\
\lVert f_{\beta, n} \rVert_1 &=& 1,\\
\lVert f_{\beta, n} \rVert_{\beta} &=& 1.
\end{eqnarray}
Hence, the Banach-Alaoglu theorem and the reflexivity of $\mathfrak{L}^{\beta}$ for $\beta \in (1,\infty)$ give us a function $f_{\beta} \in \mathfrak{L}^{\beta}$ such that some subsequence of $\{ f_{\beta, n} \}_{n=1}^{\infty}$ converges weakly in $\mathfrak{L}^{\beta}$ to $f_{\beta}$.  Without loss of generality, we assume that we have already extracted this subsequence.  Standard arguments concerning lower semi-continuity and weak convergence (c.f. \cite{FL07}) show that
\begin{eqnarray}
\mathcal{E}_p^u(f_{\beta}) &\le& 1,\\
\lVert f_{\beta} \rVert_1 &\le& 1,\\
\lVert f_{\beta} \rVert_{\beta} &\le& 1,
\end{eqnarray}
so that $f_{\beta} \in \Omega_{\beta}$ (this is the advantage of expanding the space of functions as showing that $f_{\beta} \in \mathfrak{P}_1 \cap \mathfrak{L}^{\beta}$ is difficult at this point).  Thus, we can conclude the following is true:
\begin{eqnarray}
\widetilde{\mathcal{C}}_{\beta} \le \widetilde{\Phi}_{\beta}(f_{\beta}) &=  \left( \frac{\mathcal{E}_p^u(f_{\beta})}{-\mathcal{E}_q(f_{\beta})}\right)^{3(1-\frac{1}{\beta})}\lVert f_{\beta} \rVert_{\beta} \; \lVert f_{\beta} \rVert_1^{2-\frac{3}{\beta}}&\!\!\le\!\! \left(-\mathcal{E}_q(f_{\beta})\right)^{-3(1-\frac{1}{\beta})}\!\!. \label{minInEq}
\end{eqnarray}

Since by construction \begin{equation}\lim_{n \to \infty}\widetilde{\Phi}_{\beta}(f_{\beta, n}) = \lim_{n \to \infty}\left(-\mathcal{E}_q(f_{\beta,n})\right)^{-3(1-\frac{1}{\beta})} = \widetilde{\mathcal{C}}_{\beta}, \end{equation} we will be done if we can show that $\mathcal{E}_q(f_{\beta,n})$ converges to $\mathcal{E}_q(f_{\beta})$.  Unfortunately, $\mathcal{E}_q$ is upper semi-continuous with respect to weak convergence, and so all we can immediately conclude is that \begin{equation}\mathcal{E}_q(f_{\beta}) \ge \lim_{n \to \infty}\mathcal{E}_q(f_{\beta,n}).\end{equation}

To show the convergence of $\mathcal{E}_q(f_{\beta,n})$ to $\mathcal{E}_q(f_{\beta})$, we first rewrite the potential energy functional as \begin{equation}\mathcal{E}_q(f) =-\frac{1}{2} \int \rho_f(q) \;K_f(q)\;d^3q,\end{equation} where \begin{equation}\rho_f(q) \equiv \int f(p,q) \;d^3p ,\end{equation} and \begin{equation}K_f(q) \equiv \rho_f * \lvert \textrm{Id} \rvert^{-1}(q) = \int \frac{\rho_f(q')}{\lvert q - q' \rvert}\;d^3q'. \end{equation}  We seek bounds on $\rho_{f_{\beta, n}}$ which will then imply bounds on $K_{f_{\beta, n}}$.  To that end, Lemma 4.3 in \cite{KTZ08} implies that\begin{equation}\lVert \rho_f \rVert_{\gamma_{\beta}} \le C(\beta)\lVert f \rVert_{\beta}^{\eta_{\beta}}\mathcal{E}_p^u(f)^{1-\eta_{\beta}}  \end{equation} with exponents given by \begin{equation} \gamma_{\beta} \equiv \frac{4\beta-3}{3\beta-2} \;\;\textrm{ and }\;\; \eta_{\beta} \equiv \frac{\beta}{4\beta-3}.  \end{equation}  Note that $\gamma_{\beta}$ is an increasing function of $\beta$ and that the limiting case of $\beta = \frac{3}{2}$ gives $\gamma_{\frac{3}{2}} = \frac{6}{5}$. Thus $\rho_{f_{\beta}}$ and $\rho_{f_{\beta, n}}$ are in $\mathfrak{L}^1(\mathbb{R}^3)\cap \mathfrak{L}^{\gamma_{\beta}}(\mathbb{R}^3)$ for all $\beta$ and all $n$.  Also note that the sequence $\{ \rho_{f_{\beta, n}}\}_{n=1}^{\infty}$ is uniformly bounded in $\mathfrak{L}^{\gamma_{\beta}}$-norm, and so some subsequence must converge weakly in this space.  Standard arguments show that this weak limit must equal $\rho_{f_{\beta}}$ a.e.

Next, we can conclude that $K_{f_{\beta, n}}$ and $K_{f_{\beta}}$ are in $\mathfrak{L}_{\textrm{loc}}^{\alpha}(\mathbb{R}^3)$ for $3 \le \alpha \le \frac{12\beta-9}{\beta}$ (\cite[Theorem 10.2]{LL01}).  Note that $\frac{12\beta-9}{\beta}$ is also an increasing function of $\beta$ and the limiting case of $\beta = \frac{3}{2}$ makes this exponent equal to 6.  We can turn the local estimates into global ones for all $\alpha > 3$ via the following  growth estimate on the potential.

For any spherically symmetric $f(p,q)$, the marginal mass distribution $\rho_f(q)$ will also be spherically symmetric.  Hence, the well known formula for the potential of a spherically symmetric mass distribution gives
\begin{eqnarray}
K_f(\lvert q \rvert) &=& \frac{4\pi}{\lvert q \rvert} \int_0^{\lvert q \rvert} \rho_f(r)\;r^2\;dr + 4\pi \int_{\lvert q \rvert}^{\infty} \rho_f(r)\;r\;dr\\
&\le&\frac{4\pi}{\lvert q \rvert} \int_0^{\lvert q \rvert} \rho_f(r)\;r^2\;dr + \frac{4\pi}{\lvert q \rvert} \int_{\lvert q \rvert}^{\infty} \rho_f(r)\;r^2\;dr\\
&\le&\frac{1}{\lvert q \rvert},\label{kb_bnd}
\end{eqnarray}
where in the last step we have used that for our purposes $\lVert \rho_f \rVert_1 \le 1.$  So, if $K_f \in \mathfrak{L}_{\textrm{loc}}^{\alpha}(\mathbb{R}^3)$ for $\alpha > 3$, then we have for any $R>0$
\begin{eqnarray}
\int K_f^{\alpha}(q)\;d^3q &=& \int_{\lvert q \rvert < R} K_f^{\alpha}(q)\;d^3q + \int_{\lvert q \rvert \ge R} K_f^{\alpha}(q)\;d^3q \\
&\le& \int_{\lvert q \rvert < R} K_f^{\alpha}(q)\;d^3q + \frac{4\pi}{(\alpha - 3)R^{\alpha - 3}}.
\end{eqnarray}
Thus, we may conclude that for the distributions we are considering, $K_{f_{\beta, n}}$ and $K_{f_{\beta}}$ are in $\mathfrak{L}^{\alpha}(\mathbb{R}^3)$ for $3 < \alpha \le \frac{12\beta-9}{\beta}$.  Note that this proves that $\rho_{f}K_{f} $ is indeed in $\mathfrak{L}^1(\mathbb{R}^3)$ for spherically symmetric $f \in \Omega_{\beta}$.

As with $\rho_{f_{\beta, n}}$, we have a bound on the $\mathfrak{L}^{\alpha}$-norm of $K_{f_{\beta, n}}$ that is independent of $n$, and so some subsequence of $\{K_{f_{\beta, n}}\}_{n=1}^{\infty}$ must converge weakly in this space.  As usual, this weak limit must equal $K_{f_{\beta}}$ a.e.

We can actually say something much stronger about the convergence of $\{K_{f_{\beta, n}}\}$ to $K_{f_{\beta}}$ - namely that any subsequence converging weakly to $K_{f_{\beta}}$ will actually converge strongly in $\mathfrak{L}^r$ on sets of finite measure for any $r < \frac{12\beta - 9}{\beta}$.  To see this, we note that since $\nabla \cdot \nabla K_{f_{\beta, n}}$ is in $\mathfrak{L}^1\cap\mathfrak{L}^{\gamma_{\beta}}$, we know that $\nabla K_{f_{\beta, n}}$ is locally in $\mathfrak{L}^{\frac{3}{2}} \cap \mathfrak{L}^{\kappa_{\beta}}$ where $\kappa_{\beta} = \frac{12\beta - 9}{5\beta - 3}.$  Thus, $\nabla K_{f_{\beta, n}}$ is locally in $\mathfrak{L}^2$ whenever $\beta > \frac{3}{2}.$  Arguments like the one leading to (\ref{kb_bnd}) for $K_{f_{\beta, n}}$ show that these functions decay rapidly enough at infinity to be in $\mathfrak{L}^2$ proper, and that the decay is independent of $n$.  Again, some subsequence of $\{\nabla K_{f_{\beta, n}}\}$ will converge weakly in $\mathfrak{L}^2$ and must converge to $\nabla K_{f_{\beta}}$.  Following the proof of \cite[Theorem 8.6]{LL01}, we have the strong convergence stated above.

Combining our upper bound for $K_f$ and the local strong convergence gives us that $\{K_{f_{\beta, n}}\}$ converges strongly to $K_{f_{\beta}}$ in $\mathfrak{L}^{\alpha}$ for $3 < \alpha < \frac{12\beta-9}{\beta}$:
\begin{eqnarray}
\lVert \! K_{f_{\beta, n}} \!-\! K_{f_{\beta}} \!\rVert_{\alpha}^{\alpha} &=& \lVert \left(\!K_{f_{\beta, n}} - K_{f_{\beta}}\right)\chi_{B_R(0)} \!\rVert_{\alpha}^{\alpha} \!+\! \int_{B_R^c(0)}\!\!\lvert K_{f_{\beta, n}} - K_{f_{\beta}}\!\rvert^{\alpha}d^3q\\
&\le& \lVert \left(K_{f_{\beta, n}} - K_{f_{\beta}}\right)\chi_{B_R(0)} \rVert_{\alpha}^{\alpha} + \int_{B_R^c(0)}\frac{1}{\lvert q \rvert^{\alpha}}\;d^3q\\
&\le& \lVert \left(K_{f_{\beta, n}} - K_{f_{\beta}}\right)\chi_{B_R(0)} \rVert_{\alpha}^{\alpha} + \frac{4\pi}{(\alpha -3)R^{\alpha-3}}.
\end{eqnarray}

Finally, we are in a position to show the convergence of $\mathcal{E}_q(f_{\beta,n})$ to $\mathcal{E}_q(f_{\beta})$.  We assume that $\beta > \frac{3}{2}$ and hence $\rho_f \in \mathfrak{L}^{\frac{6}{5}}$ and $K_f \in \mathfrak{L}^6$.
\begin{eqnarray}
2\lvert \mathcal{E}_q(f_{\beta}) \!-\! \mathcal{E}_q(f_{\beta, n})\!\rvert &=& \!\left|\int \rho_{f_{\beta, n}}(q)K_{f_{\beta,n}}(q) - \rho_{f_{\beta}}(q)K_{f_{\beta}}(q)d^3q\right|\\
&\le&\!\left|\int \rho_{f_{\beta, n}}(q)(K_{f_{\beta,n}}(q) - K_{f_{\beta}}(q))d^3q\right|\nonumber\\
&&\;\;\;\;\;+\!\left|\int (\rho_{f_{\beta, n}}(q) - \rho_{f_{\beta}}(q))K_{f_{\beta}}(q)d^3q\right|\\
&\le& \!\lVert\rho_{f_{\beta, n}}\rVert_{\frac{6}{5}}\lVert K_{f_{\beta, n}} \!-\! K_{f_{\beta}} \rVert_6 \!\nonumber \\
&&\;\;\;\;\;+\!\left|\int (\rho_{f_{\beta, n}}(q) - \rho_{f_{\beta}}(q))K_{f_{\beta}}(q)d^3q\right|.
\end{eqnarray}
The first term in the last inequality can be made arbitrarily small since $K_{f_{\beta, n}}$ converges strongly to $K_{f_{\beta}}$ in $\mathfrak{L}^6$ for $\beta > \frac{3}{2}$ and the $\mathfrak{L}^{\frac{6}{5}}$-norm of $\rho_{f_{\beta, n}}$ is bounded above independently of $n$ by standard interpolation estimates.  The second term can be made arbitrarily small by the weak convergence of $\rho_{f_{\beta, n}}$ to $\rho_{f_{\beta}}$ in $\mathfrak{L}^{\frac{6}{5}}$ (after possibly another subsequence extraction) and the fact that $K_{f_{\beta}}$ is in the dual space - $\mathfrak{L}^6$.

We see that inequality (\ref{minInEq}) is saturated,  and we have that \begin{equation} \widetilde{\mathcal{C}}_{\beta} = \widetilde{\Phi}_{\beta}(f_{\beta}).\end{equation}  Note that this forces
\begin{eqnarray}
\mathcal{E}_p^u(f_{\beta}) &=& 1,\\
\lVert f_{\beta} \rVert_1 &=& 1,\\
\lVert f_{\beta} \rVert_{\beta} &=& 1.
\end{eqnarray}
Thus, we have shown the existence of minimizers for $\widetilde{\Phi}_{\beta}$ over the space $\Omega_{\beta}$ satisfying the properties given above.

In the next section, we show that $f_{\beta}$ can be chosen so that it is compactly supported for $\beta > \frac{3}{2}$, so that among all possible minimizers for $\widetilde{\Phi}_{\beta}$ there is indeed one in $\mathfrak{P}_1 \cap \mathfrak{L}^{\beta}$ - proving that $\widetilde{\mathcal{C}}_{\beta} = \mathcal{C}_{\beta}$.

\section{Identification of Minimizers}

We want to find the infimum of $\widetilde{\Phi}_{\beta}$ over the space of functions $\Omega_{\beta}$ introduced in the last section.  Following an idea in Lieb-Simon (\cite{LS77}) we first note that $\Omega_{\beta}$ is a convex space of functions so that, if $f_{\beta}$ is a minimizer of our functional over $\Omega_{\beta}$ and $\eta$ is any function in this space, then for all $0 \le t \le 1$ we have that $(1-t)f_{\beta} + t\eta \in \Omega_{\beta} $.  Consequently, we can consider \begin{equation}\left.\frac{d}{dt}\right\vert_{t = 0^+} \widetilde{\Phi}_{\beta}((1-t)f_{\beta} + t\eta) , \end{equation} where by $t = 0^+$ we have in mind the one-sided Gateaux derivative from the right at zero.  This technique avoids the difficulty that arbitrary variations of a given $f\in\Omega_{\beta}$ may become negative (and hence no longer belong to $\Omega_{\beta}$).   Direct calculation gives us that this derivative is
\begin{eqnarray}
&\widetilde{\Phi}_{\beta}(f_{\beta})\left\{\left( 3-\frac{3}{\beta} \right)\left( \frac{\left.\frac{d}{dt}\right\vert_{t=0^+}\mathcal{E}_p^u((1-t)f_{\beta} + t\eta)}{\mathcal{E}_p^u(f_{\beta})} -  \frac{\left.\frac{d}{dt}\right\vert_{t=0^+}\mathcal{E}_q((1-t)f_{\beta} + t\eta)}{\mathcal{E}_q(f_{\beta})}\right)\right.& \nonumber \\
&\;\;\;\;\;\;\;\;\;\;\;\;\;\;\;\;\;\;\;\;\;\;\;\left. + \frac{\left.\frac{d}{dt}\right\vert_{t=0^+}\lVert (1-t)f_{\beta} + t\eta \rVert_{\beta}}{\lVert f_{\beta} \rVert_{\beta}} + (2-\frac{3}{\beta})\frac{\left.\frac{d}{dt}\right\vert_{t=0^+}\lVert (1-t)f_{\beta} + t\eta \rVert_1}{\lVert f_{\beta} \rVert_1} \right\}.&
\end{eqnarray}
We now compute the indicated derivatives separately.  We begin with the ultra-relativistic kinetic energy:
\begin{eqnarray}
\left.\frac{d}{dt}\right\vert_{t=0^+}\mathcal{E}_p^u((1-t)f_{\beta}+t\eta)&=&-\mathcal{E}_p^u(f_{\beta})+\iint \lvert p \rvert\; \eta(p,q)  d^3p  d^3q.
\end{eqnarray}
Next, we easily compute:
\begin{align}
\left.\frac{d}{dt}\right\vert_{t=0^+}&\!\!\lVert (1-t)f_{\beta} + t\eta \rVert_{r} = \nonumber \\
 &\lVert f_{\beta} \rVert_{r}^{1-r}\left(-\lVert f_{\beta} \rVert_{r}^r+\iint \left( f_{\beta}(p,q) \right)^{r-1}\;\eta(p,q)\;\;d^3p \; d^3q\right).
 \end{align}
Finally, we find:
\begin{eqnarray}
\left.\frac{d}{dt}\right\vert_{t=0^+}\!\!\mathcal{E}_q((1-t)f_{\beta}+t\eta) &=& \!-2\left(\mathcal{E}_q( f_{\beta}) \right) \!-\!  \iint \!\!K_{\beta}(q)\eta(p,q)d^3pd^3q
\end{eqnarray}
where we have used $K_{\beta}$ as in the previous section.  Inserting these into our derivative above, collecting terms and noting that all the constant terms cancel (i.e. those terms not involving an integration against $\eta$) yields
\begin{align}
&\left.\frac{d}{dt}\right\vert_{t = 0^+} \widetilde{\Phi}_{\beta}((1-t)f_{\beta} + t\eta)=\\
  &\iint \left[\left( 3 \! - \!\frac{3}{\beta} \right)\left(\frac{\lvert p \rvert}{\mathcal{E}_p^u(f_{\beta})} \! - \! \frac{K_{\beta}(q)}{-\mathcal{E}_q(f_{\beta})} \right)
\! + \! \frac{(f_{\beta}(p,q))^{\beta - 1}}{\lVert f_{\beta} \rVert_{\beta}^{\beta}} \! + \! \frac{2 \! - \! \frac{3}{\beta}}{\lVert f_{\beta} \rVert_1}  \right] \eta(p,q)\;d^3p\;d^3q.\nonumber
\end{align}
Since the indicator function of any set of finite measure is in $\Omega_{\beta}$, we are tempted to simply conclude that
\begin{equation} \left( 3-\frac{3}{\beta} \right)\left(\frac{\lvert p \rvert}{\mathcal{E}_p^u(f_{\beta})} - \frac{K_{\beta}(q)}{-\mathcal{E}_q(f_{\beta})} \right)
 + \frac{(f_{\beta}(p,q))^{\beta - 1}}{\lVert f_{\beta} \rVert_{\beta}^{\beta}} + \frac{2-\frac{3}{\beta}}{\lVert f_{\beta} \rVert_1}   \equiv 0\end{equation}
 or equivalently that
 \begin{equation}f_{\beta}(p,q) = \lVert f_{\beta} \rVert_{\beta}^{\frac{\beta}{\beta-1}}\left(\left( 3-\frac{3}{\beta} \right)\left( \frac{K_{\beta}(q)}{-\mathcal{E}_q(f_{\beta})} - \frac{\lvert p \rvert}{\mathcal{E}_p^u(f_{\beta})} \right) -   \frac{2-\frac{3}{\beta}}{\lVert f_{\beta} \rVert_1} \right)^{\frac{1}{\beta-1}}  .\end{equation}
Such a function cannot be in $\Omega_{\beta}$ since whenever \begin{equation}\left( 3-\frac{3}{\beta} \right)\left(\frac{K_{\beta}(q)}{-\mathcal{E}_q(f_{\beta})} - \frac{\lvert p \rvert}{\mathcal{E}_p^u(f_{\beta})} \right) -   \frac{2-\frac{3}{\beta}}{\lVert f_{\beta} \rVert_1} < 0 , \end{equation} we get complex values for $f_{\beta}$ in general.  Hence, we take
\begin{eqnarray}
f_{\beta}(p,q) &\equiv& \!\! \lVert f_{\beta} \rVert_{\beta}^{\frac{\beta}{\beta-1}}\left(\left( 3-\frac{3}{\beta} \right)\left( \!\frac{K_{\beta}(q)}{-\mathcal{E}_q(f_{\beta})} \!-\! \frac{\lvert p \rvert}{\mathcal{E}_p^u(f_{\beta})} \right) \!-\!   \frac{2-\frac{3}{\beta}}{\lVert f_{\beta} \rVert_1} \right)^{\frac{1}{\beta-1}}_+,  \label{minID}
\end{eqnarray}
where $(\cdot)_+$ means the positive part of the argument.

Since we have altered the natural minimizer so that it lies in $\Omega_{\beta}$, we need to examine the effect on the one-sided Gateaux derivatives of our functional.  Let $\Lambda$ be the support of our minimizer $f_{\beta}$.  Every $\eta \in \Omega_{\beta}$ can be decomposed as $\eta = \eta\chi_{\Lambda} + \eta\chi_{\Lambda^c}$.  Hence,
\begin{align}
&\left.\frac{d}{dt}\right\vert_{t = 0^+} \widetilde{\Phi}_{\beta}((1-t)f_{\beta} + t\eta) =\\
 &\iint_{\Lambda^c} \! \left[\left( 3 \! - \!\frac{3}{\beta} \right)\left(\frac{\lvert p \rvert}{\mathcal{E}_p^u(f_{\beta})} \! - \! \frac{K_{\beta}(q)}{-\mathcal{E}_q(f_{\beta})} \right)
\! + \! \frac{(f_{\beta}(p,q))^{\beta - 1}}{\lVert f_{\beta} \rVert_{\beta}^{\beta}} \! + \! \frac{2 \! - \! \frac{3}{\beta}}{\lVert f_{\beta} \rVert_1}  \right] \eta(p,q)\;d^3p\;d^3q.\nonumber
\end{align}
But on the the set $\Lambda^c$, we have that $f_{\beta} \equiv 0$ and that the integrand is strictly positive.  Hence, for any $\eta \in \Omega_{\beta}$
\begin{equation} \left.\frac{d}{dt}\right\vert_{t = 0^+} \widetilde{\Phi}_{\beta}((1-t)f_{\beta} + t\eta) \ge 0, \end{equation} showing that $f_{\beta}$ as defined in (\ref{minID}) is indeed a minimizer for $\widetilde{\Phi}_{\beta}$ over $\Omega_{\beta}.$

We now use scaling invariance to pick out a special minimizer that we will use to calculate $\widetilde{\mathcal{C}}_{\beta}$.  To that end, recall that the triple family of scalings as given in (\ref{scaling}) $$f_{\kappa, \lambda, \mu}(p,q) \equiv \mu \;f(\lambda p, \kappa q)  $$ (where $\kappa, \lambda$ and $\mu$ are positive real numbers) leaves $\widetilde{\Phi}_{\beta}$ invariant.  As noted in the previous section, using this scaling we can choose $f_{\beta}$ so that its $\mathfrak{L}^1$-norm is $1$.  We break from the remaining choices above by requiring $-\mathcal{E}_q(f_{\beta}) = \mathcal{E}_p^u(f_{\beta})$ (as opposed to $\mathcal{E}_p^u(f_{\beta}) = 1$).  Hence, we can no longer choose the $\mathfrak{L}^{\beta}$-norm to be $1$.  Such choices are completely arbitrary since we have an infinite number of minimizers from which to pick.  However, these choices give a minimizer of particularly nice form.

Our choices so far force \begin{equation}\kappa \lambda = \lVert f_{\beta} \rVert_1\frac{\mathcal{E}_p^u(f_{\beta})}{-\mathcal{E}_q(f_{\beta}) },\end{equation} and \begin{equation}\mu = \frac{(\kappa \lambda)^3}{\lVert f_{\beta} \rVert_1} = \lVert f_{\beta} \rVert_1^2\left(\frac{\mathcal{E}_p^u(f_{\beta})}{-\mathcal{E}_q(f_{\beta}) } \right)^3,\end{equation}  which leaves us some room to choose either $\kappa$ or $\lambda$ in a convenient way.  Using (\ref{minID}), we see that
\begin{align}
&f_{\beta; \; \kappa, \lambda, \mu}(p,q) \! = \!  \mu\left(\left( 3\! - \!\frac{3}{\beta} \right)\lVert f_{\beta} \rVert_{\beta}^{\beta}\left( \frac{K_{\beta}(\kappa q)}{-\mathcal{E}_q(f_{\beta})} \! - \! \frac{\lvert \lambda p \rvert}{\mathcal{E}_p^u(f_{\beta})} \right) \! - \!   \frac{\left(2-\frac{3}{\beta}\right)\lVert f_{\beta} \rVert_{\beta}^{\beta}}{\lVert f_{\beta} \rVert_1} \right)^{\frac{1}{\beta-1}}_+\\
&=\! \mu \! \left(\!\left(\! 3\! - \!\frac{3}{\beta} \! \right)\lVert f_{\beta} \rVert_{\beta}^{\beta}\left( \frac{\lVert f_{\beta}\rVert_1}{ -\mathcal{E}_q(f_{\beta}) } \frac{1}{\kappa} K_{\beta}^{ \kappa,\lambda,\mu}(q) \! - \! \frac{\lambda}{\mathcal{E}_p^u(f_{\beta})}\lvert p \rvert \right)\! - \!   \frac{\left(2 \! - \! \frac{3}{\beta}\right)\lVert f_{\beta} \rVert_{\beta}^{\beta}}{\lVert f_{\beta} \rVert_1} \!\right)^{\frac{1}{\beta\!-\! 1}}_+
\end{align}
where we have used $K_{\beta}^{\kappa,\lambda,\mu}(q) \equiv \int\!\!\int \frac{f_{\beta; \; \kappa,\lambda,\mu}(p',q')}{\lvert q - q'\rvert} \;d^3p'\;d^3q'$.  If we choose for $\kappa$ and $\lambda$:
\begin{equation}\kappa = \left(3 - \frac{3}{\beta} \right)\lVert f_{\beta} \rVert_1^{2\beta-1}\lVert f_{\beta} \rVert_{\beta}^{\beta}\frac{\mathcal{E}_p^u(f_{\beta})^{3\beta-3}}{(-\mathcal{E}_q(f_{\beta}))^{3\beta-2}},\end{equation}
\begin{equation}\lambda = \left(3 - \frac{3}{\beta} \right)^{-1}\frac{\lVert f_{\beta} \rVert_1^{2-2\beta}}{\lVert f_{\beta} \rVert_{\beta}^{\beta}}\frac{(-\mathcal{E}_q(f_{\beta}))^{3\beta-3}}{\mathcal{E}_p^u(f_{\beta})^{^{3\beta-4}}},\end{equation}
then we get the following:
\begin{eqnarray}
f_{\beta; \; \kappa, \lambda, \mu}(p,q) &=& \left(  \phi_{\beta; \;\kappa, \lambda, \mu}(q)-\lvert p \rvert   \right)^{\frac{1}{\beta-1}}_+,
\end{eqnarray}
where \begin{equation}\phi_{\beta; \; \kappa, \lambda, \mu}(q) \equiv   \iint \frac{f_{\beta; \; \kappa, \lambda, \mu}(p',q')}{\lvert q - q'\rvert} \;d^3p'\;d^3q' - \varkappa,\end{equation} and where the constant $\varkappa$ can be determined from the constraint \begin{equation}\lVert f_{\beta; \; \kappa, \lambda, \mu} \rVert_1 = 1.\end{equation}  From this point forward, we drop the subscripts $\kappa,\lambda,$ and $\mu$ as we now focus our attention on this particular minimizer.

Next, we determine $\phi_{\beta}$ in more detail.  We begin by computing the marginal mass distribution over configuration space:
\begin{eqnarray}
\rho_{\beta}(q) &\equiv& \int f_{\beta}(p,q) \;\;d^3p,\\
&=& 4\pi \int_{0}^{\phi_{\beta}(q)_+} \left(\phi_{\beta}(q) - \lvert p \rvert\right)^{\frac{1}{\beta - 1}} \; \lvert p \rvert^2 d\lvert p \rvert,\\
&=& \frac{8 \pi (\beta-1)^3}{\beta(2\beta - 1)(3\beta - 2)} \left( \phi_{\beta}(q) \right)^{\frac{3\beta - 2}{\beta - 1}}_+,\label{rhoeq}
\end{eqnarray}
where the last line follows by successive integration by parts.  By definition,
\begin{eqnarray}
\phi_{\beta}(q) &\equiv& \int \int \frac{f_{\beta}(p',q')}{\lvert q - q' \rvert}\;\;d^3p'\;d^3q' - \varkappa\\
&=& \frac{8 \pi (\beta-1)^3}{\beta(2\beta - 1)(3\beta - 2)} \int \frac{\left( \phi_{\beta}(q') \right)^{\frac{3\beta - 2}{\beta - 1}}_+}{\lvert q - q' \rvert}\;d^3q' -\varkappa ;
\end{eqnarray}
so upon taking the negative Laplacian of both sides we have
\begin{eqnarray}\label{minPDE}
-\triangle \phi_{\beta}(q) &=& c(\beta) \left( \phi_{\beta}(q) \right)^{\frac{3\beta-2}{\beta-1}}_+
\end{eqnarray}
with \begin{equation} c(\beta) \equiv \frac{32 \pi^2 (\beta-1)^3}{\beta(2\beta - 1)(3\beta - 2)}. \end{equation}  Our previous arguments with spherically symmetric equi-measurable rearrangements show that we can assume $\phi_{\beta}$ is spherically symmetric (where we take the center of the distribution to be the origin of our coordinate system).  We write \begin{equation}\phi_{\beta}(q) = \phi_{\beta}(\lvert q \rvert) = \phi_{\beta}(r).\end{equation} Let $R_{\beta}$ be the first zero of $\phi_{\beta}(r)$.  Then on the ball $\lvert q \rvert \le R_{\beta}$ the partial differential equation above becomes
\begin{eqnarray}
\frac{d^2 \phi_{\beta}}{dr^2} + \frac{2}{r}\frac{d \phi_{\beta}}{dr} + c(\beta)\left(\phi_{\beta } \right)^{\frac{3\beta-2}{\beta-1}}_+ &=& 0,\nonumber \\
\frac{d \phi_{\beta}}{dr}(0) &=& 0 ,\label{minODE}\\
\frac{d \phi_{\beta}}{dr} (R_{\beta}) &=& -\frac{1}{R_{\beta}^2}.\nonumber
\end{eqnarray}
Note that the first boundary condition is forced by the differential equation if we are to have a finite solution at the origin.  The second boundary condition will ensure that $f_{\beta}$ integrates to one (which also specifies the constant $\varkappa$ which we have conveniently absorbed into the definition of $\phi_{\beta}$).

Noting that the equation (\ref{minPDE}) essentially gives the gravitational potential of our mass distribution (up to a sign difference),  outside the ball $\lvert q \rvert \le R_{\beta}$  we must have \begin{equation}\phi_{\beta}(q) = \frac{1}{\lvert q \rvert} - \frac{1}{R_{\beta}}.\end{equation}  We note that as a potential function, $\phi_{\beta}$ is zero at the boundary of the mass distribution and not at infinity (as is usually the case).  This fact is easy to forget and can be the source of many headaches!

Since for each $\beta > \frac{3}{2}$ the associated density $f_{\beta}$ has compact support (c.f. section 5), we can conclude that $f_{\beta} \in \mathfrak{P}_1 \cap \mathfrak{L}^{\beta}$ as promised in the previous section.  Hence, we conclude that $\mathcal{C}_{\beta}$ (the infimum over $f_{\beta} \in \mathfrak{P}_1 \cap \mathfrak{L}^{\beta}$) is equal to the infimum over the larger space $\Omega_{\beta}$, and both are given by $\Phi_{\beta}(f_{\beta})$.  Note that this is also true for the critical case $\beta=\frac{3}{2}$ as shown in \cite{KTZ08}.

\section{Calculation of $\mathcal{C}_{\beta}$}

We begin by recalling the calculation above (c.f. (\ref{rhoeq})) for the the spatial mass density associated to $f_{\beta}$ (that is, the marginal distribution over $q$-space):
$$\rho_{\beta}(q) = \frac{8\pi(\beta-1)^3}{\beta(2\beta-1)(3\beta-2)}\left(\phi_{\beta}(q)\right)^{\frac{3\beta-2}{\beta - 1}}_+.$$

\subsection{A Most Useful Identity} \label{ident}

We begin with a simple integration by parts on the defining PDE for $\phi_{\beta}$:
\begin{eqnarray}
c(\beta) \int_{B_{R_{\beta}}(0)}\left( \phi_{\beta}(q) \right)^{\frac{4\beta-3}{\beta-1}} \;d^3q &=& \int_{B_{R_{\beta}}(0)} \lvert \nabla \phi_{\beta}(q) \rvert^2 \;d^3q,
\end{eqnarray}
since $\phi_{\beta}(q) = 0$ when $\lvert q \rvert = R_{\beta}$.

We pair this with the Pohozaev identity:
 \begin{equation} \int_{B_{R_{\beta}}(0)} \lvert \nabla \phi_{\beta}(q) \rvert^2 \;d^3q  =  \frac{6\beta-6}{4\beta-3}c(\beta)\int_{B_{R_{\beta}}(0)}\left( \phi_{\beta}(q) \right)^{\frac{4\beta-3}{\beta-1}} \;d^3q - \frac{4\pi}{R_{\beta}} .\end{equation}
This identity can be seen by first noting
\begin{equation}c(\beta)\! \int_{B_{R_{\beta}}(0)}\!\!\!\left(q \cdot \nabla \phi_{\beta}(q) \right)\left( \phi_{\beta}(q) \right)^{\frac{3\beta-2}{\beta-1}} d^3q  = \int_{B_{R_{\beta}}(0)} \!\!\! \left(q \cdot \nabla \phi_{\beta}(q) \right)\left(-\triangle \phi_{\beta}(q)\right) d^3q.\end{equation}
The left-hand integral is fairly easy:
\begin{equation} \!\int_{B_{R_{\beta}}(0)}\!\!\!\left(q \cdot \nabla \phi_{\beta}(q) \right)\left( \phi_{\beta}(q) \right)^{\frac{3\beta-2}{\beta-1}} d^3q \!=\!\! \frac{-3\beta+3}{4\beta-3}\int_{B_{R_{\beta}}(0)}\!\!\!\left( \phi_{\beta}(q) \right)^{\frac{4\beta-3}{\beta-1}} d^3q.\end{equation}
The right-hand is more involved, and after a lengthy calculation yields:
\begin{equation}\int_{B_{R_{\beta}}(0)} \left(q \cdot \nabla \phi_{\beta}(q) \right)\left(-\triangle \phi_{\beta}(q)\right) \;d^3q\!=\!-\frac{1}{2}\int_{B_{R_{\beta}}(0)} \lvert \nabla \phi_{\beta}(q) \rvert^2 \;d^3q - \frac{2\pi}{R_{\beta}}.\end{equation}
Finally, combining these two expressions for the Dirichlet integral gives
\begin{eqnarray}
\int_{B_{R_{\beta}}(0)} \lvert \nabla \phi_{\beta}(q) \rvert^2 \;d^3q &=& \frac{4\pi}{R_{\beta}}\left(\frac{4\beta-3}{2\beta-3}\right),
\end{eqnarray}
or equivalently
\begin{eqnarray}
c(\beta) \int\left( \phi_{\beta}(q) \right)^{\frac{4\beta-3}{\beta-1}}_+ \;d^3q &=& \frac{4\pi}{R_{\beta}}\left(\frac{4\beta-3}{2\beta-3}\right).
\end{eqnarray}
As we show below, this identity makes computation of the functionals comprising $\Phi_{\beta}$ remarkably easy!

\subsection{The $\mathfrak{L}^{\beta}$ Norm}

\begin{eqnarray}
\lVert f_{\beta} \rVert_{\beta}^{\beta} &=& \iint \left( f_{\beta}(p,q)\right)^{\beta} \;d^3p \;d^3q\\
&=& 4\pi \iint_{0}^{\phi_{\beta}(q)_+} \left(\phi_{\beta}(q)_+ - \lvert p \rvert\right)^{\frac{\beta}{\beta - 1}}\;\lvert p \rvert^2  \;d\lvert p \rvert \;d^3q\\
&=& \frac{8\pi(\beta-1)}{2\beta-1}\iint_{0}^{\phi_{\beta}(q)_+} \left(\phi_{\beta}(q)_+ - \lvert p \rvert\right)^{\frac{2\beta-1}{\beta - 1}}\;\lvert p \rvert  \;d\lvert p \rvert \;d^3q\\
&=&\frac{8\pi(\beta-1)^2}{(2\beta-1)(3\beta - 2)}\iint_{0}^{\phi_{\beta}(q)_+} \left(\phi_{\beta}(q)_+ - \lvert p \rvert\right)^{\frac{3\beta-2}{\beta - 1}} \;d\lvert p \rvert \;d^3q\\
&=& \frac{8\pi(\beta-1)^3}{(2\beta-1)(3\beta - 2)(4\beta - 3)}\int\left(\phi_{\beta}(q)_+\right)^{\frac{4\beta-3}{\beta - 1}} \;d^3q\\
&=& \frac{1}{R_{\beta}}\left( \frac{\beta}{2\beta - 3} \right).
\end{eqnarray}

\subsection{The Ultrarelativistic Kinetic Energy}

\begin{eqnarray}
\mathcal{E}_p^u(f_{\beta}) &=& \iint \lvert p \rvert f_{\beta}(p,q)  \;d^3p \; d^3q\\
&=& 4\pi\iint_0^{\phi_{\beta}(q)_+} \left(\phi_{\beta}(q)_+ - \lvert p \rvert\right)^{\frac{1}{\beta - 1}}\;\lvert p \rvert^3 \;d\lvert p \rvert \;d^3q\\
&=& \frac{12\pi(\beta - 1)}{\beta}\iint_0^{\phi_{\beta}(q)_+} \left(\phi_{\beta}(q)_+ - \lvert p \rvert\right)^{\frac{\beta}{\beta - 1}}\;\lvert p \rvert^2 \;d\lvert p \rvert \;d^3q\\
&=& \frac{24\pi(\beta - 1)^2}{\beta(2\beta-1)}\iint_0^{\phi_{\beta}(q)_+} \left(\phi_{\beta}(q)_+ - \lvert p \rvert\right)^{\frac{2\beta-1}{\beta - 1}}\;\lvert p \rvert \;d\lvert p \rvert \;d^3q\\
&=& \frac{24\pi(\beta - 1)^3}{\beta(2\beta-1)(3\beta-2)}\iint_0^{\phi_{\beta}(q)_+} \left(\phi_{\beta}(q)_+ - \lvert p \rvert\right)^{\frac{3\beta-2}{\beta - 1}} \;d\lvert p \rvert \;d^3q\\
&=&\frac{24\pi(\beta - 1)^4}{\beta(2\beta-1)(3\beta-2)(4\beta-3)}\int \left(\phi_{\beta}(q)_+ \right)^{\frac{4\beta-3}{\beta - 1}}  \;d^3q\\
&=& \frac{3(\beta-1)}{4\pi(4\beta-3)}c(\beta)\int \left(\phi_{\beta}(q)_+ \right)^{\frac{4\beta-3}{\beta - 1}}  \;d^3q\\
&=& \frac{3\beta-3}{R_{\beta}(2\beta-3)}.
\end{eqnarray}

\subsection{The Potential Energy}

Our choice of scaling is meant to ensure that the potential energy of our minimizers equals their ultra-relativistic kinetic energy.  To check this, we calculate the potential energy directly.
\begin{eqnarray}
-\mathcal{E}_q(f_{\beta}) &=&\frac{1}{2} \iiiint \frac{f_{\beta}(p',q')f_{\beta}(p,q)}{\lvert q - q' \rvert} \;d^3p' \; d^3p \; d^3q' \; d^3q\\
&=& \frac{1}{2}\iint \frac{\rho_{\beta}(q')\rho_{\beta}(q)}{\lvert q - q' \rvert}\; d^3q' \; d^3q\\
&=& -\frac{(\beta-1)^3}{\beta(2\beta-1)(3\beta-2)}\iint \frac{\left( \phi_{\beta}(q)_+ \right)^{\frac{3\beta-2}{\beta-1}}\triangle_{q'} \phi_{\beta}(q')}{\lvert q - q' \rvert}\; d^3q' \; d^3q\\
&=&\!\! \frac{(\beta\!-\!1)^3}{\beta(2\beta\!-\!1)(3\beta\!-\!2)}\int_{B_{R_{\beta}}(0)} \!\!\left( \phi_{\beta}(q) \right)^{\frac{3\beta-2}{\beta-1}}\left[\int_{B_{R_{\beta}}(0)} \!\!\frac{-\triangle_{q'} \phi_{\beta}(q')}{\lvert q \!-\! q' \rvert} d^3q' \right] d^3q.\nonumber\\
\end{eqnarray}
We first work on the bracketed integral before tackling the entire expression:
\begin{eqnarray}
\int_{B_{R_{\beta}}(0)} \frac{-\triangle_{q'} \phi_{\beta}(q')}{\lvert q - q' \rvert}\; d^3q' &=& \int_{B_{R_{\beta}}(0)} \nabla_{q'} \phi_{\beta}(q') \cdot \nabla_{q'} \left(\frac{1}{\lvert q - q' \rvert}\right)\; d^3q'\nonumber\\&& - \int_{\partial B_{R_{\beta}}(0)}\left(\frac{1}{\lvert q - q' \rvert}\right) \nabla_{q'} \phi_{\beta}(q') \cdot \;d\vec{\sigma'}\\
&=& \int_{B_{R_{\beta}}(0)} \phi_{\beta}(q')  \triangle_{q'} \left(\frac{-1}{\lvert q - q' \rvert}\right)\; d^3q'\nonumber\\
&& +\int_{\partial B_{R_{\beta}}(0)} \phi_{\beta}(q')\;\nabla_{q'} \left(\frac{1}{\lvert q - q' \rvert}\right) \cdot \;d\vec{\sigma'}\nonumber\\&& - \int_{\partial B_{R_{\beta}}(0)}\left(\frac{1}{\lvert q - q' \rvert}\right) \nabla_{q'} \phi_{\beta}(q') \cdot \;d\vec{\sigma'}\\
&=& 4\pi \left(\phi_{\beta}(q) + \frac{1}{R_{\beta}}\right).
\end{eqnarray}
Inserting this expression into the functional above gives:
\begin{eqnarray}
-\mathcal{E}_q(f_{\beta}) &=&\frac{4\pi(\beta\!-\!1)^3}{\beta(2\beta\!-\!1)(3\beta\!-\!2)}\int_{B_{R_{\beta}}(0)} \!\!\left( \phi_{\beta}(q) \right)^{\frac{3\beta\!-\!2}{\beta\!-\!1}}\left[ \phi_{\beta}(q) \!+\! \frac{1}{R_{\beta}}\right] d^3q\\
&=&\frac{c(\beta)}{8\pi}\int_{B_{R_{\beta}}(0)} \left( \phi_{\beta}(q) \right)^{\frac{4\beta-3}{\beta-1}}\; d^3q + \frac{1}{2R_{\beta}}\int_{B_{R_{\beta}}(0)} \rho_{\beta}(q)\; d^3q\\
&=& \frac{1}{2R_{\beta}}\left(\frac{4\beta-3}{2\beta-3}\right)+\frac{1}{2R_{\beta}}\\
&=& \frac{3\beta-3}{R_{\beta}(2\beta-3)}.
\end{eqnarray}

\subsection{The Formula for $\mathcal{C}_{\beta}$}

We see from our work above that the ultra-relativistic kinetic energy and potential energy are indeed equal for $f_{\beta}$ (as we chose in our scaling).  Thus, the only contribution to $\mathcal{C}_{\beta}$ comes from the $\mathfrak{L}^{\beta}$ norm.  Hence, we arrive at the following formula:
\begin{eqnarray}
\mathcal{C}_{\beta} &=& \left( \frac{\beta}{R_{\beta}(2\beta - 3)} \right)^{\frac{1}{\beta}}.
\end{eqnarray}

\section{$\mathcal{C}_{\beta}$ and the Standard Lane-Emden Polytropes}

Though the formula for $\mathcal{C}_{\beta}$ given above is rather elegant, $R_{\beta}$ is only defined implicity by our requirement that $\lVert f_{\beta} \rVert_{1} = 1$. In order to compute $\mathcal{C}_{\beta}$ we rewrite its formula in terms of the solutions of the famous \emph{Lane-Emden ODE} with standard initial data.  In the usual notation (as in \cite{Ch67}) this ODE is
\begin{eqnarray}
\frac{d^2\theta_n}{d\xi^2} + \frac{2}{\xi}\frac{d\theta_n}{d\xi} + \theta_n^n &=& 0, \nonumber\\
\theta_n(0) &=& 1,\\
\frac{d\theta_n}{d\xi}(0) &=& 0 \nonumber.
\end{eqnarray}
The solution to this ODE for a particular choice of $n$ is often referred to as \emph{the standard polytrope of index $n$}.  It is well-known (c.f. \cite{Ch67}) that the standard polytropes for $n \in [0,5)$ first cross the $\xi$-axis at a finite distance from the origin.  This first zero is often denoted $\xi_n$.

Explicit solutions are only known for three indices:
\begin{eqnarray}
\theta_0(\xi) &=& 1-\frac{\xi^2}{6},\\
\theta_1(\xi) &=& \frac{\sin(\xi)}{\xi},\\
\theta_5(\xi) &=& \frac{1}{\sqrt{1+\frac{1}{3}\xi^2}},
\end{eqnarray}
giving $\xi_0 = \sqrt{6}$, $\xi_1 = \pi$, and $\xi_5 = \infty$.  Of equal importance is the slope of $\theta_n$ at the first zero.  In the cases above, we have:
\begin{eqnarray}
\frac{d\theta_0}{d\xi}(\xi_0) &=& -\frac{\sqrt{6}}{3},\\
\frac{d\theta_1}{d\xi}(\xi_1) &=& -\frac{1}{\pi},\\
\lim_{\xi \to \infty}\frac{d\theta_5}{d\xi}(\xi) &=& 0.
\end{eqnarray}

We next explore rescaling the standard polytropes in order to find functions which satisfy our equation for $\phi_{\beta}$ (\ref{minODE}).  We first note that the polytropic indices arising in the determination of $\mathcal{C}_{\beta}$ range over $(3,5]$ (so that $n=1$ is clearly avoided in our considerations).  We make the following definition (for $n \ne 1$) \begin{equation} \gamma_n(\xi) \equiv \alpha_n^{-1} A_n^{\frac{2}{n-1}}\theta_n(A_n\xi),\end{equation} and note the following consequences:
\begin{eqnarray}
\gamma_n\left( \frac{\xi_n}{A_n} \right) &=& 0,\\
\frac{d\gamma_n}{d\xi}(\xi) &=& \alpha_n^{-1} A_n^{\frac{n+1}{n-1}}\frac{d\theta_n}{d\xi}(A_n\xi),\\
\frac{d^2\gamma_n}{d\xi^2}(\xi) &=& \alpha_n^{-1} A_n^{\frac{2n}{n-1}}\frac{d^2\theta_n}{d\xi^2}(A_n\xi).
\end{eqnarray}
These change our ODE to
\begin{eqnarray}
\alpha_n A_n^{\frac{-2n}{n-1}} \frac{d^2\gamma_n}{d\xi^2} + \frac{2}{A_n\xi}\alpha_nA_n^{-\frac{n+1}{n-1}}\frac{d\gamma_n}{d\xi} + \alpha_n^n A_n^{\frac{-2n}{n-1}}\gamma_n^n &=& 0,
\end{eqnarray}
which reduces to
\begin{eqnarray}
\frac{d^2\gamma_n}{d\xi^2} + \frac{2}{\xi}\frac{d\gamma_n}{d\xi} + \alpha_n^{n-1} \gamma_n^n &=& 0,\\
\gamma_n(0) &=& \alpha_n^{-1} A_n^{\frac{2}{n-1}},\\
\frac{d\gamma_n}{d\xi}(0) &=& 0.
\end{eqnarray}

We clearly must have \begin{equation} n(\beta) = \frac{3\beta-2}{\beta-1} ,\end{equation} where $n$ runs from $5$ down to $3$ as $\beta$ runs from $\frac{3}{2}$ up to infinity.  Equivalently, we have \begin{equation} \beta(n)  = \frac{n-2}{n-3}.\end{equation}  It is also clear that we will need \begin{equation} \alpha_{n(\beta)} = c(\beta) ^{\frac{1}{n(\beta)-1}}.\end{equation}

The determination of $A_{n(\beta)}$ comes from the second boundary condition of (\ref{minODE}):
\begin{eqnarray}
\frac{d\gamma_{n(\beta)}}{d\xi}\left(\frac{\xi_{n(\beta)}}{A_{n(\beta)}}\right) &=& \alpha_{n(\beta)}^{-1} A_{n(\beta)}^{\frac{n(\beta)+1}{n(\beta)-1}}\frac{d\theta_{n(\beta)}}{d\xi}(\xi_{n(\beta)})\\
&=& -\frac{A_{n(\beta)}^2}{\xi_{n(\beta)}^2}.
\end{eqnarray}
Some algebra reveals that \begin{equation} A_{n(\beta)} = \left( \frac{-\xi_{n(\beta)}^2 \theta'_{n(\beta)}(\xi_{n(\beta)})}{\alpha_{n(\beta)}} \right)^{\beta} ,\end{equation} which leads to the following formula for $R_{\beta}$ in terms of the standard Lane-Emden data:  \begin{equation}R_{\beta} = \xi_{n(\beta)}\left( -\xi_{n(\beta)}^2\theta'_{n(\beta)}(\xi_{n(\beta)})\right)^{1-2\beta}(c(\beta))^{\beta-1}. \end{equation}
This in turn leads to the following very useful (but decidedly more cumbersome) formula for $\mathcal{C}_{\beta}$: \begin{equation} \mathcal{C}_{\beta} = \left( \frac{\beta(-\xi_{n(\beta)}^2 \theta'_{n(\beta)}(\xi_{n(\beta)}))^{2\beta-1}}{(2\beta-3)\xi_{n(\beta)}(c(\beta))^{\beta-1}} \right)^{\frac{1}{\beta}} .\end{equation}  Incidentally, we have the following formula for $\phi_{\beta}:$ \begin{equation*} \phi_{\beta}(q) = \alpha_{n(\beta)}^{-1}A_{n(\beta)}^{\frac{2}{n(\beta)-1}}\theta_{n(\beta)}(A_{n(\beta)}q), \end{equation*}  which we do not expand further for reasons of brevity!
\section{Numerical Results}

It will be beneficial to rewrite our formula for $\mathcal{C}_{\beta}$ in terms of the standard polytropic index $n(\beta)$:
\begin{eqnarray}
\mathcal{C}_{\beta} &=& \left( \frac{(n(\beta)-2)(-\xi_{n(\beta)}^2 \theta'_{n(\beta)}(\xi_{n(\beta)}))^{\frac{n(\beta)-1}{n(\beta)-3}}}{(5-n(\beta))\xi_{n(\beta)}(c(\beta))^{\frac{1}{n(\beta)-3}}} \right)^{\frac{n(\beta)-3}{n(\beta)-2}}.
\end{eqnarray}
We first compute the value of $\mathcal{C}_{\beta}$ as $\beta = 3/2$.

First, recall the following facts (\cite{Bu78}):
\begin{eqnarray}
\lim_{n \to 5} -\xi_{n}^2 \theta'_{n}(\xi_{n}) &=& \sqrt{3},\\
\lim_{n \to 5} (5-n)\xi_n &=& \frac{32\sqrt{3}}{\pi}.
\end{eqnarray}
Since $n(3/2) = 5$ and $c(3/2) = 8\pi^2/15$, we see that \begin{equation}\mathcal{C}_{\frac{3}{2}} = \frac{3}{8} \left( \frac{15}{16} \right)^{\frac{1}{3}},  \end{equation}
which reproduces the value found in \cite{KTZ08} despite the fact that our analysis does not apply to the limiting case of non-compactly supported minimizers.  Note that the polytrope of index 5 (commonly referred to as the Plummer Sphere in the astrophysical literature) is not compactly supported.

Using Maple to run the numerical approximations for the standard polytropes yields the following plot of $\mathcal{C}_{\beta}$ (displayed with the bounds found in \cite{KTZ08} and a vertical line at $\beta=3/2$ indicating that $\mathcal{C}_{\beta}=0$ for $1 < \beta < 3/2$):
\begin{figure}[ht]\centering
  \includegraphics[height=125mm,width=135mm]{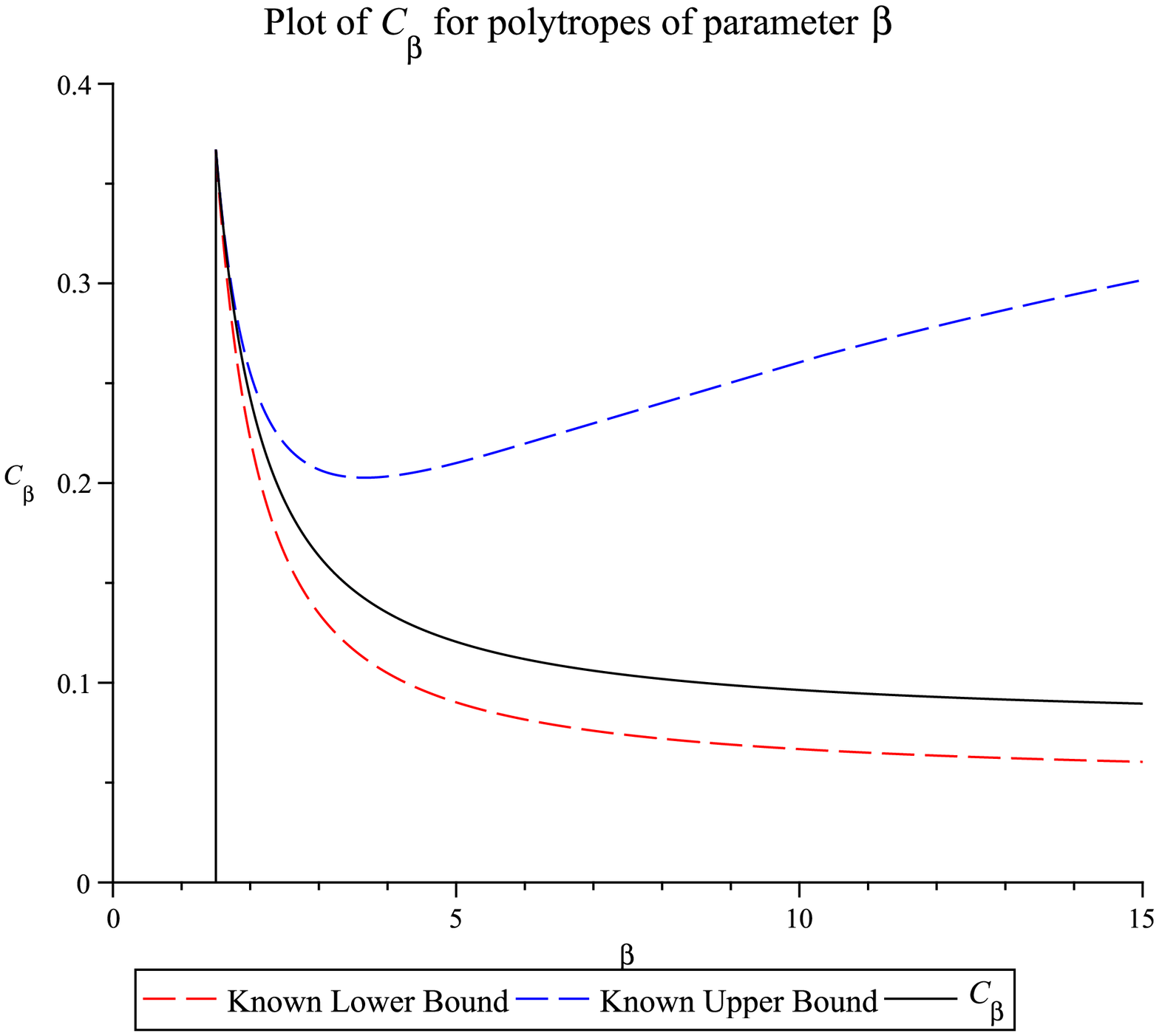}\\
  \label{fig1}
\end{figure}

\newpage
Incidentally, we can also give an improved upper bound over that listed in \cite{KTZ08}.  To see this, note that the upper bound in (\ref{lb_ub}) is convex decreasing between
$\beta=3/2$ and $\beta \approx 3.6649$ beyond which it is strictly increasing for all $\beta$  and converges to a finite value (namely $45/8\pi^2$) as $\beta$ tends to infinity.  In contrast,  the numerical evaluation of $\mathcal{C}_{\beta}$ is a decreasing function of $\beta$.  Hence we can improve the upper bound given in (\ref{lb_ub}) by simply replacing it with its convex hull.  We find (using Maple to estimate the minimum) that the improved upper bound takes the constant value $0.20269$  for $\beta \ge 3.6649$.

\section{Asymptotics}

For the limiting behavior as $\beta$ tends to infinity, we need that \begin{equation} \lim_{\beta \to \infty} c(\beta) = \frac{16\pi^2}{3}. \end{equation}
We also note that since $\lim_{\beta \to \infty} n(\beta) = 3$, the only terms that contribute are:  \begin{equation} \mathcal{C}_{\infty} =  \frac{3(-\xi_{3}^2 \theta'_{3}(\xi_{3}))^{2}}{16\pi^2}.  \end{equation}  This exact expression is a little less than illuminating since $\theta_3$ is not known explicitly.  However, there are extensive numerical data available.  Referring to \cite[p. 407]{Ho86}, we see that $-\xi_{3}^2 \theta'_{3}(\xi_{3}) \approx 2.018236.$  Thus, we can at least conclude that \begin{equation} \mathcal{C}_{\infty} \approx 0.077383. \end{equation}  In comparison, Theorem I of \cite{GS85}  (recalling that the total mass is 1 in our considerations) requires that initial data have $\mathcal{L}^{\infty}$-norm less than $40^{-3} \approx 0.00002.$  Of course, Glassey and Schaeffer did not aim for the optimal constant and so were generous in their estimates.  In comparison, the lower bound given in \cite{KTZ08} is approximately $0.049438.$

Since the standard polytrope of index $5$ is known explicitly, we can find an asymptotic expression for $\mathcal{C}_{\beta}$ when $\beta$ is sufficiently close to $3/2$.  We begin by examining an identity involving the zeroes of the standard polytrope of index $n$: \begin{equation} \frac{n+1}{(5-n)\xi_n} = \frac{\int_0^{\xi_n}\left( \theta_n(r) \right)^{n+1}r^2\;dr}{(-\xi_n^2\theta_n'(\xi_n))^2} ,\end{equation} (this is essentially a reformulation of the identity found in \ref{ident} for the standard polytropes).  The right-hand side limits to a finite value as $n$ approaches $5$ (\cite{Bu78}): \begin{equation} \lim_{n \to 5} \frac{\int_0^{\xi_n}\left( \theta_n(r) \right)^{n+1}r^2\;dr}{(-\xi_n^2\theta_n'(\xi_n))^2} = \frac{\pi\sqrt{3}}{16}.\end{equation}  So, for $n$ sufficiently close to $5$, we have that \begin{equation} \xi_n \approx \frac{16(n+1)}{\pi\sqrt{3}(5-n)},\end{equation} and accordingly, for $\beta$ sufficiently close to $\frac{3}{2}$ \begin{equation}R_{\beta} \approx \frac{16}{3\pi}\left( \frac{4\beta-3}{2\beta-3} \right) \left(\frac{32\pi^2(\beta-1)^3}{3\beta(2\beta-1)(3\beta-2)}  \right)^{\beta-1} .\end{equation}  Finally, this yields an asymptotic expression for $\mathcal{C}_{\beta}$ near $\frac{3}{2}$:
\begin{eqnarray}
\mathcal{C}_{\beta} &\approx& \left[\frac{3\pi}{16}\left(\frac{\beta}{4\beta-3} \right) \right]^{\frac{1}{\beta}}\left( \frac{3\beta(2\beta-1)(3\beta-2)}{32\pi^2(\beta-1)^3} \right)^{1-\frac{1}{\beta}}.
\end{eqnarray}

\begin{figure}[ht]\centering
  \includegraphics[width=90mm]{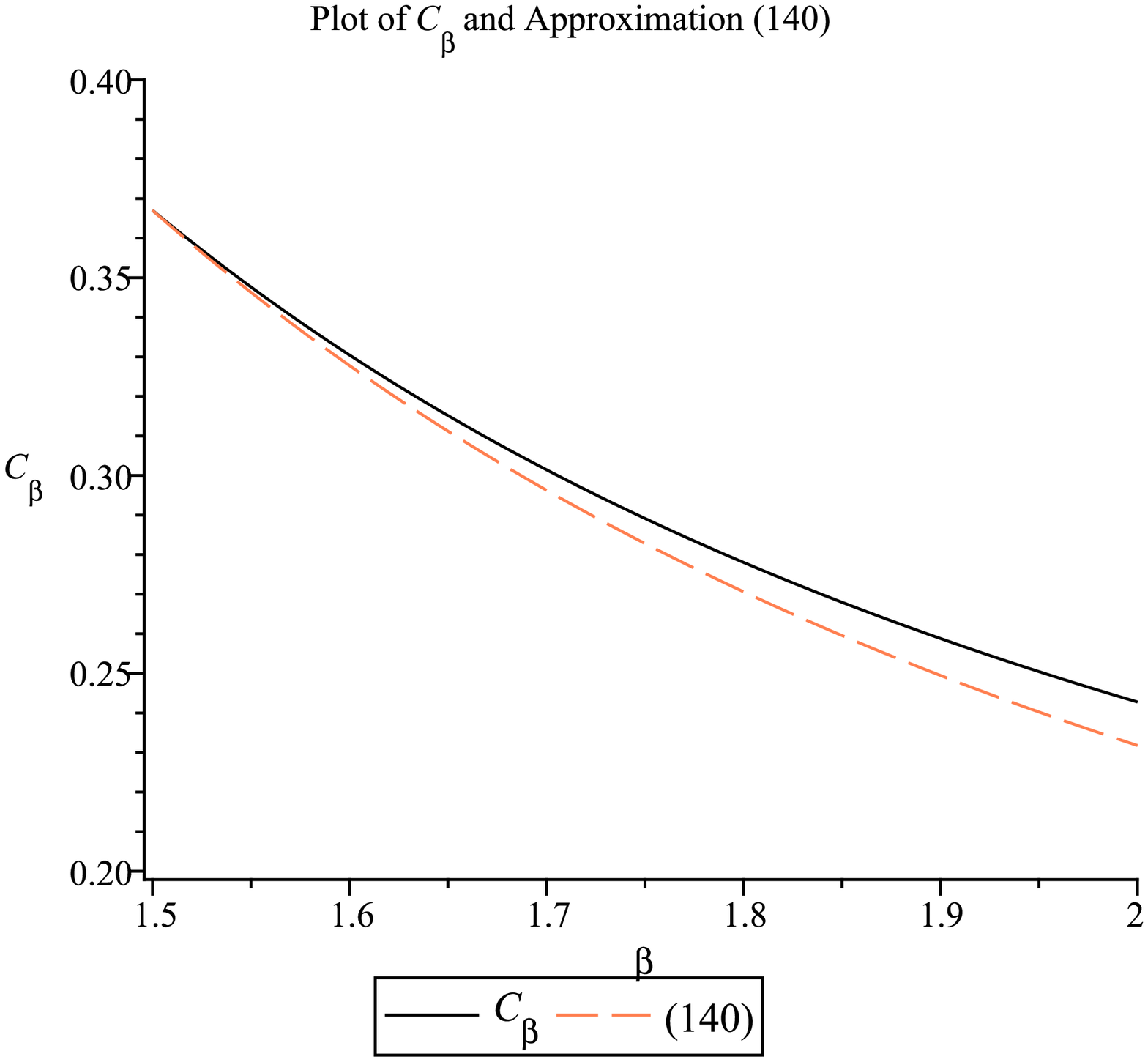}\\
  \label{fig2}
\end{figure}

\section{Acknowledgements}

This work was supported by NSF grant DMS 08-07705 to Michael Kiessling.  The author thanks Michael Kiessling for proposing the problem.  In addition, the author wishes to offer many thanks to both Michael Kiessling and Shadi Tahvildar-Zadeh for numerous useful and enlightening conversations.  Thanks go also to Yves Elskens and the other organizers of Vlasovia III for a stimulating meeting where the results of this paper were presented as a poster, and special thanks to Vlasovia III for providing supplementary financial support.  Finally, the author wishes to thank both anonymous referees for their very helpful comments on this paper.

\end{document}